\documentclass[seceq,preprint]{ptptex}
\usepackage{wrapft}

\newcommand{\be}{\begin{equation}}
\newcommand{\ee}{\end{equation}}
\newcommand{\bea}{\begin{eqnarray}}
\newcommand{\eea}{\end{eqnarray}}
\newcommand{\bean}{\begin{eqnarray*}}
\newcommand{\eean}{\end{eqnarray*}}
\newcommand{\ba}{\begin{array}}
\newcommand{\ea}{\end{array}}

\newcommand{\slashm}[1]{\ooalign{\hfil/\hfil\crcr$#1$}}

\pubinfo{Vol.~113, No.~1, January 2005}
\preprintnumber[3cm]{
KEK-TH-901\\
JUPD-0410\\
hep-ph/0410327\\
October, 2004}

\title{
Unravelling Soft Components in the Shape Function 
\\ 
for Inclusive B Decays
}

\author{
Hiroyuki \textsc{Kawamura},$^{1,}$\footnote{E-mail: hiroyuki.kawamura@kek.jp} 
Jiro \textsc{Kodaira}$^{2,}$\footnote{E-mail: kodaira@theo.phys.sci.hiroshima-u.ac.jp} 
and Kazuhiro \textsc{Tanaka}$^{3,}$\footnote{E-mail: tanakak@sakura.juntendo.ac.jp}
}

\inst{
$^1$Theory Group, KEK, Tsukuba, Ibaraki 305-0801, Japan
\\
$^2$Dept. of Physics, Hiroshima University, Higashi-Hiroshima 739-8526, Japan
\\
$^3$Dept. of Physics, Juntendo University, Inba-gun, Chiba 270-1695, Japan
}

\recdate{
October 26, 2004
}

\abst{
In the heavy-quark limit, 
the decay spectrum for radiative and semileptonic inclusive $B$-meson decays 
is determined by a universal structure function, i.e. the shape function.
We study the constraints on the bilocal heavy-quark operator for the shape function 
from 
the QCD equations of motion and 
heavy-quark symmetry,
and we obtain a new basis of nonlocal operators that reveals
relevant soft components in the shape function.
Those  
nonlocal operators 
represent 
the ``kinetic energy distribution'' of the $b$-quark inside the $B$-meson 
and the four-parton correlations with additional quarks and gluons.
A corresponding local operator basis 
relevant for relating 
the shape function to HQET parameters is also given,
and novel effects of the quark-gluon
correlations are discussed.
}

\begin{document}

\maketitle

\section{Introduction}

Inclusive $B$-meson decays, such as charmless semileptonic 
$B \rightarrow X_{u} \ell \bar{\nu}$ decays
and penguin-induced $B \rightarrow X_{s}\gamma$ decays, 
are of special interest because they are sensitive probes
of electroweak parameters as well as new physics.
Theoretical analysis of the corresponding (differential) decay rates 
has been developed on the basis of the operator product expansion (OPE) 
as a power series in $\Lambda_{\rm QCD}/m_{b}$~\cite{Chay:1990da}.
This
provides justification for the parton model
and also the power corrections to it 
in terms of 
matrix elements of local operators~\cite{Bigi:1993fe}.
It is known, however, that the usual OPE becomes singular 
near the kinematic endpoint $E/E_{\rm max} \sim 1$
for the lepton (photon) energy spectrum 
in the decay $B \rightarrow X_{u} \ell \bar{\nu}$ 
($B \rightarrow X_{s}\gamma$);
the endpoint region plays an important role in experimental analysis,
especially for the precise determination of the CKM matrix element $|V_{ub}|$
from $B \rightarrow X_{u} \ell \bar{\nu}$,
avoiding large backgrounds from the decays into charmed particles,
$B \rightarrow X_{c} \ell \bar{\nu}$.
In the endpoint region, the hadronic decay products 
evolving from the $u$-quark ($s$-quark)
in $B \rightarrow X_{u} \ell \bar{\nu}$ 
($B \rightarrow X_{s}\gamma$)
have large energy but small invariant mass, so that
the hadronic final state becomes jet-like with collinear interactions of an outgoing light quark.
This implies that 
the short-distance expansion is not applicable, receiving the light-cone singularity,
and 
the OPE has to be reorganized
with the resummation of the most singular terms~\cite{Neubert:1993um,Bigi:1993ex}.
This is accomplished through the light-cone expansion, 
which is similar to   
the treatment in deep inelastic lepton-nucleon scattering (DIS).
As the result, 
at the leading power of $\Lambda_{\rm QCD}/m_{b}$,
the shape of the decay spectrum 
is described by an analogue of the leading twist in the DIS, 
i.e., by the factorization formula,
where a 
structure function 
corresponding to the nonperturbative, 
long-distance ($\sim 1/\Lambda_{\rm QCD}$) contribution
is convoluted with the perturbatively calculable 
function~\cite{Neubert:1993um,Bigi:1993ex,Korchemsky:1994jb}.
The structure function, called the shape function,
is process independent and universal in the sense that
the same shape function determines the decay spectrum of both 
$B \rightarrow X_{u} \ell \bar{\nu}$ 
and
$B \rightarrow X_{s}\gamma$.
More rigorous 
treatments of the factorization 
with three relevant mass scales, hard ($m_b$), hard-collinear ($\sqrt{m_{b}\Lambda_{\rm QCD}}$), and 
soft ($\Lambda_{\rm QCD}$), 
have been presented recently 
in the framework of soft-collinear effective theory (SCET)
\cite{Bauer:2001yt,Bauer:2003pi,blnp,ne}.
Those treatments have shown that 
the perturbatively
calculable function is 
composed of 
two parts, the hard function 
due to the hard corrections
and the jet function due to hard-collinear fluctuations
associated with the light-quark jet,
and that
the resulting
factorization formulae for $B \rightarrow X_{u} \ell \bar{\nu}$ 
and
$B \rightarrow X_{s}\gamma$
are valid to all orders in $\alpha_{s}$~\cite{Korchemsky:1994jb,Bauer:2001yt,Bauer:2003pi,blnp}.

One-loop corrections~\cite{Corbo:1982ah}
as well as the resummation of 
large (leading and next-to-leading) endpoint 
logarithms~\cite{Al,Korchemsky:1994jb,Akhoury:1995fp,Grozin:1994ni,Bauer:2003pi,blnp,ne}
have been calculated for perturbatively calculable functions in the factorization formulae
for $B \rightarrow X_{u} \ell \bar{\nu}$ 
and
$B \rightarrow X_{s}\gamma$.
Also for the shape function,
one-loop renormalization group evolution 
has been studied by many authors~\cite{Balzereit:1998yf,Aglietti:1999ur,Grozin:1994ni,Bauer:2003pi,blnp,ne}.
Unfortunately, however, our understanding of the nonperturbative boundary value 
in the corresponding evolution equation is still poor.
There have been some attempts to constrain the boundary value of the shape function
by estimating the relevant nonperturbative effects,
e.g., trying to relate the moments of the shape function with the fundamental parameters
in the heavy-quark effective theory (HQET)~\cite{Bigi:1993ex,Neubert:1993um,Mannel:1994pm,Bigi:it}, 
and examining the ambiguity (``infrared renormalons'') of the perturbation series 
for the shape function~\cite{Korchemsky:1994jb,Grozin:1994ni}.
Guided by such constraints, one would construct ans\"{a}tze for the shape function, and 
eventually fit them
to experimental data to fix the remaining uncertainties~\cite{Kagan:1998ym,Chen:2001fj}.

In attempting to reach this goal, 
it is desirable to analyze 
the operator structure of the shape function in QCD and
clarify the maximal model-independent constraints among the
matrix elements of the relevant operators.
However, a systematic study for this purpose has not yet been carried out. 
In this paper, 
we discuss a systematic operator analysis 
for the shape function.  
Our approach is based on the exact identity for the bilocal heavy-quark operator,
which we derive from the QCD equations of motion and heavy-quark symmetry.
Combined with the light-cone expansion of the bilocal operator 
to separate the longitudinal-momentum dependence from transverse one,
it is possible to derive a differential equation that controls the longitudinal-momentum dependence of 
the shape function.
We find that the corresponding differential equation involves the ``source terms'' given by
matrix elements of 
a finite set of 
higher-dimensional nonlocal operators.
Those  
nonlocal operators 
represent 
the ``kinetic energy distribution'' of the $b$-quark inside the $B$-meson 
and the four-parton correlations with additional quarks and gluons.
Solving this differential equation,
we express the shape function in terms of 
those nonlocal operators that explicitly represent the
nonperturbative effects relevant for the shape function.

When deriving the differential equation from the exact operator identity, and also when solving the differential equation,
we need to specify the boundary conditions for the shape function.
In this paper, we employ the boundary conditions that the (non-negative) moments of the shape function 
as a function of the longitudinal momentum 
are finite. 
These conditions are equivalent to assuming that the nonlocal operators relevant to the shape function
are the generating functions of the corresponding local operators via the Taylor expansion,
and hold trivially when ignoring renormalization effects.\footnote{It is well known 
that similar conditions are satisfied by the structure functions in the DIS, and that they
are preserved by including the radiative corrections~\cite{Balitsky:1987bk,Balitsky:1990ck,Ali:1991em,Kodaira:1998jn}.}
However, recent results for the one-loop renormalization of the shape function in the $\overline{\rm MS}$ scheme
claim that the renormalization effects completely modify the mathematical properties as well as the 
physical interpretation of the shape function. In particular, a 
``radiative tail'' is generated for the momentum representation of the shape function, so that  
all non-negative moments become ultra-violet (UV) divergent~\cite{Balzereit:1998yf,Grozin:1994ni,Bauer:2003pi,blnp}.
Because of this, the differential equation for the shape function and its solution discussed in this paper
are exact up to the strong renormalization effects of order $\alpha_{s}$.
However, it is important to emphasize that actually our results are useful and indispensable beyond the lowest order in the 
perturbative effects, 
i.e., for obtaining the total behavior of the renormalized shape function: As noted in 
{Refs.~\citen{Grozin:1994ni,Bauer:2003pi,blnp}},
the strong UV behavior in the renormalization effects 
implies that, for a consistent treatment and interpretation, the renormalized shape function should be 
further ``factorized'' into the ``hard components'' involving the radiative corrections and 
the ``soft components'' involving all the nonperturbative effects.
In the matching calculation to perform this factorization, the first step is to list 
a basis of operators to describe the relevant soft components. 
The solution given in the present paper provides
a basis of nonlocal operators as well as the corresponding local operator basis,
including the four-parton correlation operators that were previously unknown.
Furthermore, our solution provides the complete tree-level result for the hard components.

The remainder of this paper is organized as follows.
In \S2 we demonstrate that the equations of motion 
and heavy-quark symmetry
allow us to express the bilocal heavy-quark operator for the shape function in terms of 
a finite set of 
higher-dimensional nonlocal operators,
which provides a new basis of nonlocal operators 
representing
relevant soft components in the shape function.
We also derive a corresponding local operator basis in \S3, which is 
relevant when relating 
the soft components of the shape function to the HQET parameters.
In contrast to the previous 
works~\cite{Bigi:1993ex,Neubert:1993um,Mannel:1994pm,Bigi:it,Korchemsky:1994jb,Grozin:1994ni}, 
our local operator basis is composed of 
an increasing number of operators with increasing dimension; 
in addition to a few first operators with low dimension, whose matrix elements  
give the well-known HQET parameters~\cite{Neubert:1993um,Mannel:1994pm,Bigi:1993ex},
we find ``new'' operators with higher dimensions, whose matrix elements
provide the ``generalized'' HQET parameters that
represent the ``Fermi motion'' of the $b$-quark as well as the four-parton correlation effects
inside the $B$-meson.
We also discuss
simple estimates of some of these generalized HQET parameters 
and novel effects of the quark-gluon
correlations.
In \S4 we derive an explicit formula 
for the momentum representation of the shape function.
Although this formula is subject to modification
due to large radiative corrections,
it deserves consideration, because it
expresses the ``Fermi motion'' of the $b$-quark 
in an explicit analytic form and the four-parton correlation effects in 
an integral representation. 
In \S5 we present our conclusions.

\section{Nonlocal HQET operators and equations of motion}

The factorization formulae for the decay spectrum in the inclusive $B$-meson decays are proved 
with the two-step matching, i.e., matching QCD onto SCET at a scale $\sim m_{b}$,
followed by matching SCET onto HQET
at a scale $\sim \sqrt{m_{b}\Lambda_{\rm QCD}}$~\cite{Korchemsky:1994jb,Bauer:2001yt,Bauer:2003pi,blnp,ne}.
The shape function is introduced at 
the second step 
by integrating out the final-state light-quark jet associated with the mass scale $\sqrt{m_{b}\Lambda_{\rm QCD}}$,
and it is defined as 
the $B$-meson
matrix element of a gauge-invariant, bilocal light-cone operator 
in the HQET~\cite{Neubert:1993um,Bigi:1993ex,Korchemsky:1994jb,Bauer:2001yt}:
\be
\langle \bar{B}(v) | \bar{h}_{v}(z) [z,0]
h_{v}(0) |\bar{B}(v) \rangle
= \tilde{f}(t)= \int d\omega e^{i\omega t} f(\omega) \ .
\label{eq:sf0}
\ee
Here $z^{\mu}$ denotes a light-like vector, $z^{2}=0$, 
$v^{\mu}$ is the 4-velocity of the $B$-meson ($v^2 =1$),
and $t=v\cdot z$.
We choose the Lorentz frame of the system as $z^{\mu}=(0, z^{-}, \boldsymbol{0}_{\perp})$
and $v^{\mu}=(v^{+},v^{-}, \boldsymbol{0}_{\perp})$, corresponding to 
the case in which the final-state jet momentum 
points in the ``$-$'' direction on the light-cone. 
$h_{v}(x)$ denotes the effective $b$-quark field,
$b(x) \approx \exp(-im_{b} v\cdot x)h_{v}(x)$,
and it is subject to the on-shell constraint
$\slashm{v} h_{v} = h_{v}$~\cite{Isgur:1989vq,Neubert:1994mb}.
Here 
\be
[z,0] = {\rm P} \exp \left(ig\!\! \int_0^1 \!\! d\xi\, z_\mu A^\mu (\xi z)\right)
\label{eq:wl}
\ee
is the path-ordered gauge factor along the straight line connecting the points $z$ and $0$.
{}For brevity, 
in the following, we do not show
the path-ordered gauge factors 
connecting the constituent fields.
We employ 
a mass-independent normalization of the
$B$-meson state $|\bar{B}(v) \rangle$,
such that 
$\langle \bar{B}(v)|\bar{B}(v') \rangle = v^{0}(2 \pi/m_{B})^{3}\delta^{(3)}(v-v')$,
with $m_B$ the $B$-meson mass. 
The functions $\tilde{f}(t)$ and $f(\omega)$ give the coordinate
and (residual) momentum representations of the shape function, respectively.
Due to the heavy-quark spin symmetry, there appears no other independent 
function by taking matrix elements of the bilocal operators 
with other Dirac matrices inserted. 

Actually, the shape function (\ref{eq:sf0}) depends on the scale $\mu$ 
at which the nonlocal light-cone operator
is renormalized ($\sqrt{m_{b}\Lambda_{\rm QCD}} \gtrsim \mu \gtrsim \Lambda_{\rm QCD}$). 
It is known that the radiative corrections induce a Sudakov-type
strong scale dependence on (\ref{eq:sf0}),
governed by the corresponding
cusp anomalous dimension
\cite{Korchemsky:1994jb,Grozin:1994ni,Balzereit:1998yf,Aglietti:1999ur}, 
and, corresponding to this effect, $\tilde{f}(t)$ in the $\overline{\rm MS}$ scheme becomes 
singular for short distances:
$\tilde{f}(0)$, as well as all derivatives of $\tilde{f}(t)$ at $t=0$, diverge. 
As discussed in \S1, however, such singular UV behavior can be 
factorized from the ``soft components'' of the shape function by an additional matching procedure,
and we can ignore the renormalization
in order to disentangle the operator structure relevant for the soft components.
Therefore, in the following, we work with ``naive'' mathematical properties 
for (\ref{eq:sf0}) that are valid at lowest order in the perturbative effects,
e.g., that $\tilde{f}(0)=1$, and that $\tilde{f}(t)$ can be Taylor expanded about $t=0$. 
With these boundary conditions,
the physical meaning of $f(\omega)$ is that it is the distribution of the 
residual momentum $k^+=\omega v^+$ of the heavy quark inside the $B$-meson.\footnote{
Actually, the validity of this ``probabilistic interpretation'' appears to be critical,
due to nonperturbative as well as perturbative effects. (See the discussion in \S\S4 and 5 below.)}
A similar approach has been employed in Refs.~\citen{Korchemsky:1994jb} and \citen{Grozin:1994ni}
to analyze a few first local operators in the Taylor expansion
of $\bar{h}_{v}(z) h_{v}(0)$ about $z_{\mu}=0$, considering
the small $t$ expansion of $\tilde{f}(t)$.
Our approach corresponds to an extension of the analysis 
given in Refs.~\citen{Korchemsky:1994jb} and \citen{Grozin:1994ni}
by treating the relevant nonlocal operator 
$\bar{h}_{v}(z) h_{v}(0)$ directly.
Although this is formally equivalent to 
treating all local operators in the Taylor expansion
of $\bar{h}_{v}(z) h_{v}(0)$
simultaneously,
our nonlocal operator approach has the advantage of making 
the operator structure for the soft components of the shape function most transparent.

We can utilize the nonlocal operator technique, 
which has been
developed for analyzing the (higher twist) nucleon structure 
functions~\cite{Balitsky:1987bk,Balitsky:1990ck,Ali:1991em,Kodaira:1998jn}
(see also Refs.~\citen{Braun:1990iv} and \citen{Ball:1999ff}\footnote{
A similar extension 
has been employed 
for 
the $B$-meson light-cone wavefunctions~\cite{Kawamura:2001jm}. 
}).
The nonperturbative dynamics of the $b$-quark, surrounded by the light quarks, antiquarks, and
gluons,
reveals itself as the response of the nonlocal operator $\bar{h}_{v}(z) h_{v}(0)$
to the change of the interquark separation and/or total translation,
but the total translation is irrelevant for the forward matrix elements.
The relevant response of the nonlocal operator 
is described by the exact operator identity
\be 
v^{\mu}\frac{\partial}{\partial x^{\mu}}\bar{h}_{v}(x) 
h_{v}(0) = 
\bar{h}_{v}(x)
v\ \cdot\! 
\stackrel{\leftarrow}{D} h_{v}(0)
+ i \int_{0}^{1}duu \ \bar{h}_{v}(x) gG_{\mu \nu}(ux) v^{\mu} x^{\nu}
  h_{v}(0) \ ,
\label{eq:id}
\ee
where $x^{\mu}$ is not restricted on the light cone.  
$\stackrel{\leftarrow}{D}_{\mu} =
\stackrel{\leftarrow}{\partial}_{\mu} + ig A_{\mu}$
and $D_{\mu}= \partial_{\mu} - igA_{\mu}$
are the covariant derivatives, and 
$G_{\mu \nu}= (i/g)[D_{\mu}, D_{\nu}]$
is the gluon field strength tensor.
Taking the $B$-meson matrix element of the relation~(\ref{eq:id}),
the first term on the RHS vanishes, due to the HQET equations of motion 
$\bar{h}_{v}\ v\ \cdot \!\stackrel{\leftarrow}{D}\ =0$.
We take the light-cone limit, $x_{\mu} \rightarrow z_{\mu}$;
for the calculation of the LHS,
we actually need to extend the definition (\ref{eq:sf0}) to the case
in which the interquark separation is not light-like, where we have
\be
\langle \bar{B}(v) | \bar{h}_{v}(x) h_{v}(0) |\bar{B}(v) \rangle
= \tilde{f}(v\cdot x)  + x^{2}\tilde{F}(v\cdot x) + {\cal O}(x^{4})\ ,
\label{eq:sf1}
\ee
while, for the second term on the RHS,
we introduce a three-parton correlation function 
$\tilde{R}(t, u)$ as
\be
\langle \bar{B}(v) | \bar{h}_{v} (x) \, g G_{\mu\nu} (ux)\, x^{\nu}
      \,  h_{v} (0) | \bar{B}(v) \rangle
  = \frac{1}{2}\left(x_{\mu}- \frac{x^{2}}{v\cdot x}v_{\mu}\right)\left[\tilde{R}(t, u) + {\cal O}(x^{2})\right] \ .
\label{eq:3part}
\ee

Then we obtain the following constraint equation due to the equations of motion:
\be
\frac{d \tilde{f}(t)}{dt} + 2t 
\tilde{F}(t)
= \frac{i}{2}t \int_{0}^{1}  du\, u\, \tilde{R}(t, u) \ .
\label{eq:de0}
\ee
This equation is formally similar to 
the corresponding differential equations
for the twist-3 nucleon structure functions~\cite{Kodaira:1998jn}, 
but an important difference 
is the participation of $\tilde{F}(t)$ in (\ref{eq:de0}),
which expresses the effect due to 
the deviation from the light-cone, as in (\ref{eq:sf1}). 
{}For the nucleon case, contributions of this type correspond 
to twist-4 effects and decouple from the equations at the twist-3 level,
because the twist defined as ``dimension minus spin'' of the relevant operators
is a good quantum number. 
Although the twist counting of the shape function (\ref{eq:sf0})
would be twist-3, such a ``conventional'' twist
is no longer a useful concept in the HQET.

It is straightforward to see that the insertion of an arbitrary Dirac matrix $\Gamma$ into 
(\ref{eq:id})-(\ref{eq:sf1}) 
does not lead to a new equation for $\tilde{f}(t)$.
This is unlike the case for the $B$-meson wavefunctions~\cite{Kawamura:2001jm},
and it is due to so strong constraints
on the forward matrix elements imposed by
the heavy-quark spin symmetry.

To proceed, we analyze the explicit operator structure of
$\tilde{F}(t)$ in (\ref{eq:de0}), which 
can be extracted from the next-to-leading term
in the light-cone expansion 
of the nonlocal operator $\bar{h}_{v}(x) h_{v}(0)$ [see (\ref{eq:sf1})].
{}For this purpose, we utilize an elegant method
to construct the light-cone expansion~\cite{Balitsky:1987bk,Balitsky:1990ck,Ball:1999ff}:
The leading term (lt) in the light-cone expansion,
corresponding to the leading twist operator in the DIS,
obeys the equation
\be 
\frac{\partial^{2}}{\partial x_{\mu}\partial x^{\mu}}
\left[ \bar{h}_{v}(x) h_{v}(0) \right]_{\rm lt} =0\ ,
\label{eq:lap}
\ee
which ensures that all local operators arising in the Taylor expansion
are traceless. 
A formal solution is~\cite{Balitsky:1987bk}
\bea
\left[ \bar{h}_{v}(x) h_{v}(0) \right]_{\rm lt} 
&&= \bar{h}_{v}(x) h_{v}(0)
\nonumber \\
+&&
\sum_{n=1}^{\infty}
\frac{(-x^{2}/4)^{n}}{n! (n-1)!}\int_{0}^{1} du 
\frac{(1-u)^{n-1}}{u^{n}}
\left(\frac{\partial^{2}}{\partial x_{\mu} \partial x^{\mu}}
\right)^{n} 
\bar{h}_{v}(ux) h_{v}(0)\ .
\label{eq:formal}
\eea
To order $x^{2}$,
we obtain
\be
\bar{h}_{v}(x) h_{v}(0) = \left[ \bar{h}_{v}(x) h_{v}(0) \right]_{\rm lt}
+ \frac{x^{2}}{4} \int_{0}^{1}\! \frac{du}{u}\
\frac{\partial^{2}}{\partial x_{\mu} \partial x^{\mu}}
\bar{h}_{v}(ux) h_{v}(0)+ {\cal O}(x^{4})\ .
\label{eq:formal2}
\ee
Actually, the first term on the RHS of (\ref{eq:sf1})
contains the ${\cal O}(x^{2})$ term as well as the leading term.
To separate contributions of different powers in $x^{2}$,
it is convenient to 
exploit the light-cone expansion of 
$\exp(i \omega v\cdot x)$
entering into the definition of the momentum representation 
of $\tilde{f}(v\cdot x)$~\cite{Balitsky:1990ck,Ball:1999ff},
\be 
e^{i \omega v\cdot x}
= \left[ e^{i \omega v\cdot x} \right]_{\rm lt}
- \frac{\omega^{2} x^{2}}{4}\int_{0}^{1}du u e^{i u \omega v\cdot x}
+ {\cal O}(x^{4})\ ,
\label{eq:exp}
\ee
where $\left[ e^{i \omega v\cdot x} \right]_{\rm lt}$ is defined by 
a straightforward generalization of the procedure 
$\left[ \ldots \right]_{\rm lt}$ 
of (\ref{eq:formal}) to an arbitrary function of $x$.
Substituting (\ref{eq:formal2})-(\ref{eq:exp}) into (\ref{eq:sf1})
and comparing both sides of the resulting equation,
the leading terms reproduce the definition (\ref{eq:sf0}),
and we get, from the next-to-leading ${\cal O}(x^{2})$ terms, 
\be
\tilde{F}(t) =
\frac{1}{4} \int_{0}^{1}du u \tilde{\Phi}(ut)
+ \frac{1}{4} \int d\omega \omega^{2} f(\omega)
\int_{0}^{1}du u e^{i u \omega t}\ ,
\label{eq:le}
\ee
with
\be
\tilde{\Phi}(t)= 
\left.
\langle \bar{B}(v) | 
\frac{\partial^{2}}{\partial x_{\mu} \partial x^{\mu}}
\bar{h}_{v}(x) h_{v}(0) 
|\bar{B}(v) \rangle
\right|_{x\rightarrow z}\ .
\label{eq:Om}
\ee
Here, the second term on the RHS of (\ref{eq:le})
is the analogue of Nachtmann's correction in the DIS.
An important point is that this term depends on $f(\omega)$, and it produces an additional
term involving $\tilde{f}(t)$ in (\ref{eq:de0}).
Substituting (\ref{eq:le}) into (\ref{eq:de0}), 
we obtain
\be
t\frac{d \tilde{f}(t)}{dt} + \tilde{f}(t) 
-1
= t^{2}\int_{0}^{1}  du\, u\, \left( i\tilde{R}(t, u) 
- \tilde{\Phi}(ut)
\right)\ .
\label{eq:de1}
\ee
Here 
the RHS involves $\tilde{\Phi}$ of (\ref{eq:Om}), which 
is concerned with the effects of the transverse as well as 
longitudinal motion of the $b$-quark inside the $B$-mesons.
To reveal its physical content further, the following 
exact identity is useful:\footnote{
A similar identity for the case of the light quarks
is discussed in Ref.~\citen{Ball:1999ff}.}
\bea
\frac{\partial^{2}}{\partial x_{\mu} \partial x^{\mu}}
\bar{h}_{v}(x) h_{v}(0) 
&=& \bar{h}_{v}(x)(\stackrel{\leftarrow}{D}_{\mu} )^{2} h_{v}(0) 
+2i\int_{0}^{1}du u \frac{\partial}{\partial x_{\mu}}
\bar{h}_{v}(x)\, g G_{\mu\nu} (ux)\, x^{\nu}\, h_{v} (0) 
\nonumber\\
&-&i \int_{0}^{1}du u^{2} \bar{h}_{v}(x)\, \left[ D^{\mu}, g G_{\mu\nu} (ux)\right]\, 
x^{\nu}\, h_{v} (0) 
\nonumber\\
+&&\!\!\!\!\!\!  2\int_{0}^{1}du u \int_{0}^{u}ds s 
\bar{h}_{v}(x)\, g G_{\mu\nu} (ux)\, x^{\nu}\, g G^{\mu\rho} (sx)\, x_{\rho}\,h_{v} (0)\ .
\label{eq:id2}
\eea
This can be derived straightforwardly by a method similar 
to (\ref{eq:id}).
Substituting (\ref{eq:id2}) 
into 
(\ref{eq:Om}),
the replacement 
$\stackrel{\leftarrow}{D}_{\mu} \rightarrow 
\stackrel{\leftarrow}{D}_{\perp \mu}
\equiv \stackrel{\leftarrow}{D}_{\mu}-v_{\mu}v\, 
\cdot \stackrel{\leftarrow}{D}$
can be made in the first term on the RHS 
by using the equations of motion for the effective heavy-quark field.
Similarly, 
the equations of motion for the gluons imply
$\left[ D^{\mu}, G_{\mu\nu} \right]
= -gt^{a}\sum \left( v_{\nu} \bar{h}_{v}t^{a} h_{v} 
+ \bar{q}t^{a}\gamma_{\nu}q\right)$,
with the summation ($\sum$) over the heavy ($h_{v}$) and light ($q$) flavors.
Therefore, 
(\ref{eq:Om})
can be expressed in terms of a set of many-body nonlocal operators.
This pattern is similar to that for the twist-4 effects 
in the DIS~\cite{Balitsky:1987bk,Balitsky:1990ck},
but one important difference is that the covariant derivative for 
the transverse direction, ${D}_{\perp \mu}$, acting on the quark fields
cannot be completely eliminated in the present case;
participation of such a transverse derivative in 
a complete set of the higher-dimensional operators 
is typical of the HQET~\cite{Isgur:1989vq,Neubert:1994mb}. 
In fact, the corresponding operator in (\ref{eq:id2})
plays an important role in describing the Fermi motion effects 
in the $B$-mesons, as we show below. 

The matrix element of the derivative of the three-body operator in 
(\ref{eq:id2}) can be calculated 
using (\ref{eq:3part}).
An important observation is that this contribution cancels 
the first term $i\tilde{R}(t, u)$ on the RHS of (\ref{eq:de1}),
so that the three-parton correlation $\tilde{R}$ decouples from our differential equation.
For the remaining terms, we define 
\bea
\langle \bar{B}(v) | \bar{h}_{v}(z)(\stackrel{\leftarrow}{D_{\perp}}_{\mu})^{2} 
h_{v}(0)| \bar{B}(v) \rangle
=\langle \bar{B}(v) | \bar{h}_{v}(z)(\stackrel{\rightarrow}{D_{\perp}}_{\mu})^{2} 
h_{v}(0)| \bar{B}(v) \rangle
=&&  
\tilde{K}(t)\ ,
\label{eq:K}
\\
\langle \bar{B}(v)| \bar{h}_{v}(z) 
\left[ D^{\mu}, gG_{\mu \nu}(uz) \right] z^{\nu} h_{v}(0)|\bar{B}(v) \rangle
\;\;\;\;\;\;\;\;\;\;\;\;\;\;\;\;\;\;\;\;\;\;\;\;\;\;\;\;\;\;\;\;\;\;\;&&
\nonumber\\
= - \sum_{q}
\langle \bar{B}(v)| \bar{h}_{v}(z) 
g^{2}t^{a}\bar{q}(uz)t^{a}\slashm{z}q(uz) h_{v}(0)|\bar{B}(v) \rangle
=  t \tilde{W}(t, u)\ ,&& \label{eq:tw}
\\
\langle \bar{B}(v)| \bar{h}_{v}(z) 
gG_{\mu \nu}(uz) z^{\nu} gG^{\mu \rho}(sz) z_{\rho} h_{v}(0)|\bar{B}(v) \rangle
=  t^2 \tilde{Y}(t, u, s)\ . &&\label{eq:ty}
\eea
Here, $\tilde{K}(t)$ can be interpreted as the ``kinetic energy distribution'' 
of the $b$-quark inside the $B$-meson, while $\tilde{W}(t,u)$ and $\tilde{Y}(t,u,s)$ are
the four-parton correlations with additional quarks and gluons, respectively.
We denote the RHS of (\ref{eq:de1}) as $\tilde{J}(t)$ and substitute (\ref{eq:K})-(\ref{eq:ty})
into it.
We then obtain 
\begin{equation}
%
\tilde{J}(t) =\int_{0}^{t} dt' \left\{-t' \tilde{K}(t') + i{t'}^2 \int_{0}^{1}du u^2 \tilde{W}(t', u)
-2 {t'}^3 \int_{0}^{1} du u \int_{0}^{u} ds s\ \tilde{Y}(t', u, s)\right\}. \
\label{eq:jtilde}
\end{equation}
Regarding $\tilde{J}(t)$ as the ``source'' term,
(\ref{eq:de1}) is immediately integrated to give, 
with the condition $\tilde{f}(0)=1$,
\be
\tilde{f}(t) = 1 + \frac{1}{t}\int_{0}^{t} d\tau \tilde{J}(\tau)
\ .
\label{eq:sol1}
\ee
Thus $\tilde{f}(t)$ is completely re-expressed 
in terms of a set of nonperturbative functions, $\tilde{K}(t), \tilde{W}(t,u)$ and $\tilde{Y}(t,u,s)$,
corresponding to the higher-dimensional nonlocal operators.

\section{Local operator basis and HQET parameters}

In this section we derive a basis of local composite operators to represent the soft components
of the shape function, 
and we discuss the HQET parameters as matrix elements of these operators. 
It is straightforward to expand the solution~(\ref{eq:sol1}) 
in a power series in $t$ as
\begin{equation}
\tilde{f}(t) = 1+\sum_{n=1}^{\infty}\frac{(it)^{n}}{n!}{\cal A}_{n}\ ,
\label{eq:power}
\end{equation}
where the $n$-th term 
gives  
the nonperturbative power corrections 
of order $(\Lambda_{\rm QCD}t)^n$.
We obtain
\begin{eqnarray}
{\cal A}_n &=& \frac{n-1}{n+1}{\cal K}_{n-2}
- \frac{1}{n(n+1)}\left[ \sum_{k=0}^{\left[\frac{n-3}{2}\right]}
\left\{(k+1)(k+2)
\right.\right.
\nonumber\\
&+&\left. \left.(n-k-1)(n-k-2)\right\} {\cal W}_{k, n-3-k}
-\frac{1-(-1)^{n}}{2}\frac{n^2 -1}{4}{\cal W}_{\frac{n-3}{2},\frac{n-3}{2}}\right]
\nonumber\\
&-&\frac{2}{n(n+1)}\sum_{k=0}^{n-4}\left[ \sum_{l=0}^{\left[\frac{k}{2}\right]}
\left\{ (l+1)(n+l-k-1)+(k-l+1)(n-l-1)\right\}{\cal Y}^{n-4-k}_{l, k-l}\right.
\nonumber\\
&-&\left.
\frac{1+(-1)^k}{2}\left(\frac{k}{2}+1\right)\left(n-\frac{k}{2}-1\right){\cal Y}_{\frac{k}{2},\frac{k}{2}}^{n-4-k}
\right]\ ,
\label{eq:momsol}
\end{eqnarray}
for $n= 1, 2, 3, \ldots$. Here ${\cal K}_{n}, {\cal W}_{l, k}$ and ${\cal Y}_{l,k}^{m}$
denote matrix elements of the local operators, 
which are generated from the Taylor expansion of (\ref{eq:K})-(\ref{eq:ty})
about $z_{\mu}=0$:
\begin{eqnarray}
{\cal K}_{n} 
&=& \frac{1}{(v^{+})^{n}}
\langle \bar{B}(v)| \bar{h}_{v} (iD^{+})^{n} (D_{\perp \mu})^2 h_{v}|\bar{B}(v) \rangle\ ,
\label{eq:momek}\\
{\cal W}_{l, k} 
&=& -\frac{1}{(v^{+})^{k+l+1}}
\sum_{q} \langle \bar{B}(v)| \bar{h}_{v}(iD^{+})^{k}  
g^{2}t^{a}\bar{q}t^{a}\gamma^{+} q 
(iD^{+})^{l} h_{v}|\bar{B}(v) 
\rangle\ , \label{eq:momew}\\
{\cal Y}_{l,k}^{m}
&=&\frac{1}{(v^{+})^{k+l+m+2}}
\langle \bar{B}(v)| \bar{h}_{v}(iD^{+})^{k} g{G_{\mu}}^{+} (iD^{+})^{m} gG^{\mu +} (iD^{+})^{l}h_{v}|\bar{B}(v) 
\rangle\ . \label{eq:momey}
\end{eqnarray}
Note that ${\cal W}_{l, k}$ and ${\cal Y}_{l,k}^{m}$
are symmetric under the interchange $l \leftrightarrow k$, which follows from the behavior of the
corresponding matrix elements (\ref{eq:momew}) and (\ref{eq:momey}) under
the parity transformation combined with the time-reversal
transformation.
Equation (\ref{eq:momsol})
gives, for $n=1$ and 2,
\be
{\cal A}_{1}=0, \;\;\;\;\;\;\;\;\;\;\;\;\;\;\;\;\;\;\;\;\;\;\;\;\;\;\;\;\;\;
{\cal A}_{2} = \frac{1}{3}{\cal K}_{0}=- \frac{\lambda_{1}}{3}\ , 
\label{eq:mnt}
\ee
with the fundamental HQET parameter $\lambda_{1}$
related to the average kinetic energy of 
the heavy quark inside the $B$-mesons~\cite{Isgur:1989vq,Neubert:1994mb},
\be
\lambda_{1} = \langle \bar{B}(v) | \bar{h}_{v}\left(i D_{\perp \mu}\right)^{2} h_{v}
| \bar{B}(v) \rangle\ ,
\label{eq:lambda1}
\ee
and, for $n=3$,
\be
{\cal A}_{3} = - \frac{1}{6}{\cal W}_{0,0}=
\frac{1}{6}
\sum_{q} \langle \bar{B}(v)| \bar{h}_{v}
g^{2}t^{a}\bar{q}t^{a}\slashm{v} q  h_{v}|\bar{B}(v) \rangle\ , 
\label{eq:a3}
\ee
where the four-quark operator is related to the quark-gluon Darwin operator by the equations of motion 
[see also (\ref{eq:tw})].
Equation (\ref{eq:mnt}) coincides with that obtained in
Refs.~\citen{Korchemsky:1994jb} and \citen{Grozin:1994ni}. Also, (\ref{eq:a3}) as well as (\ref{eq:mnt})
coincides with the corresponding results discussed 
in the earlier works~\cite{Bigi:1993ex,Neubert:1993um,Mannel:1994pm}.
We note that those works 
analyzed the constraints resulting from the equations of motion on the matrix elements of 
the corresponding local operator,
\be
{\cal A}_{n} = \frac{1}{(v^{+})^{n}} \langle \bar{B}(v) | \bar{h}_{v}
\left(iD^{+}\right)^{n} h_{v} | \bar{B}(v) \rangle
\ ,
\label{eq:annn}
\ee
which follows from (\ref{eq:sf0}).

Furthermore, (\ref{eq:momsol}) gives new results for $n\ge 4$,
\bea
{\cal A}_{4}&=&\frac{3}{5}{\cal K}_{2} - \frac{2}{5}{\cal W}_{0,1} - \frac{3}{10}{\cal Y}_{0,0}^{0}\ ,
\label{eq:fourth}\\
{\cal A}_{5}&=&\frac{2}{3}{\cal K}_{3}- 
\frac{7}{15}{\cal W}_{0,2} - \frac{1}{5}{\cal W}_{1,1} - 
\frac{4}{15}{\cal Y}_{0,0}^{1} - \frac{11}{15}{\cal Y}_{0,1}^{0}\ ,
\label{eq:fifth}
\eea
and so forth. Therefore, the operators in (\ref{eq:momek})-(\ref{eq:momey}), whose matrix elements give 
\bea
&&{\cal K}_{n-2}, \;\;\;\;\;\;\;\; \;\;\;\; \;\;\;\; 
{\cal W}_{k, n-3-k} \;\;\;\; \left(k=0, 1, 2, \cdots, \left[\frac{n-3}{2}\right]\right)\ ,
\nonumber \\
&&\;\;\;\; {\cal Y}^{n-4-k}_{l, k-l}\;\;\;\; 
\left(k=0, 1, 2, \cdots,  n-4;\ l=0, 1, 2, \cdots,  \left[\frac{k}{2}\right]\right)\ ,
\label{eq:basis}
\eea
form a basis of local operators of dimension $n+3$ and contribute to ${\cal A}_{n}$.
This implies that for $n \gg 1$ there are 
$\sim n^{2}/4$ independent local operators participating in ${\cal A}_{n}$.

Here it is worth studying the details of the new nonperturbative
matrix elements contributing to ${\cal A}_{4}$,
because these matrix elements provide a generalization of the HQET parameters.
${\cal K}_{2}$ and ${\cal W}_{0,1}$, given by (\ref{eq:momek}) and (\ref{eq:momew}),
can be related to the ``trace part'' of the corresponding operators as
\begin{eqnarray}
{\cal K}_2 &=& \frac{1}{3}\langle \bar{B}(v)| \bar{h}_{v}  \left((D_{\perp \mu})^2\right)^2 h_{v}|\bar{B}(v) \rangle\ ,
\label{eq:momek2}\\
{\cal W}_{0, 1} &=&  
\frac{i}{3}\sum_{q} \langle \bar{B}(v)| \bar{h}_{v} g^{2}t^{a}\bar{q}t^{a}\gamma_{\mu} q 
D_{\perp}^{\mu} h_{v}|\bar{B}(v) \rangle\ . 
\label{eq:momew2}
\end{eqnarray}
Thus ${\cal K}_{2}>0$, and 
a comparison of (\ref{eq:momek2}) with (\ref{eq:lambda1}) leads to 
the rough estimate ${\cal K}_{2} \sim \lambda_{1}^{2}$.
The matrix element (\ref{eq:momew2}) involving the four-quark operator
can be evaluated in the vacuum saturation (``factorization'') approximation\footnote{
This approximation gives the estimate 
${\cal W}_{0,0}\approx \pi [(N_{c}^{2}-1)/2N_{c}^2]\alpha_{s}f_{B}^{2}m_{B}$ 
of (\ref{eq:a3})~\cite{Bigi:1993ex,Mannel:1994pm}.} 
as
\be
{\cal W}_{0,1}\approx -\pi \frac{N_{c}^{2}-1}{6N_{c}^2}\alpha_{s}f_{B}^{2}m_{B}\bar{\Lambda}\ ,
\label{eq:satu}
\ee
with the usual decay constant $f_{B}$, 
defined as 
$\langle 0|\bar{q}\gamma^{\mu}\gamma_{5}h_{v}|\bar{B}(v) \rangle=if_{B}\sqrt{m_{B}/2}\ v^{\mu}$.
Here $\bar{\Lambda}$ represents 
the asymptotic value of 
the mass difference
$m_{B}-m_{b}$
between the $B$-meson and the $b$-quark~\cite{Isgur:1989vq,Neubert:1994mb},
and it can be identified with the effective mass of the light degrees of freedom
in the $B$-meson. Using phenomenological values for the parameters on the RHS of (\ref{eq:satu}) 
leads to ${\cal W}_{0,1} \sim -\lambda_{1}^{2}$.
Contrastingly, ${\cal Y}_{0,0}^{0}$ of (\ref{eq:momey})
cannot be estimated in a simple manner, but in the $B$-meson rest frame 
it is expressed in terms of the chromoelectric and chromomagnetic fields as
\begin{equation}
{\cal Y}_{0,0}^{0}=-\frac{2}{3}
\langle \bar{B}(
{\mbox{\boldmath $v$}}
={\mbox{\boldmath $0$}}
)| \bar{h}_{v} g^{2} \left( {\mbox{\boldmath $E$}}^{2}+{\mbox{\boldmath $B$}}^{2} 
\right) h_{v}|\bar{B}({\mbox{\boldmath $v$}}
={\mbox{\boldmath $0$}}) 
\rangle\ , \label{eq:momey2}
\end{equation}
so that ${\cal Y}_{0,0}^{0} <0$.
Here we note that a comparison between (\ref{eq:mnt}) and (\ref{eq:annn}) with $n=4$ suggests
the rough estimate
${\cal A}_{4} \sim \lambda_{1}^2$, and combined with the relation (\ref{eq:fourth}),\footnote{
The identity between matrix elements of the local operators, which is implied by (\ref{eq:annn}) with $n=4$
and (\ref{eq:fourth}), would be subject to the perturbative ${\cal O}(\alpha_{s})$ corrections, but these corrections
do not affect this simple estimate.}
we would expect ${\cal Y}_{0,0}^{0}\sim - \lambda_{1}^{2}$.
In principle, ${\cal A}_{5}$ of (\ref{eq:fifth})
could be analyzed similarly, but it involves the five new nonperturbative
matrix elements and is much more complicated.

When $\Lambda_{\rm QCD} t$ is not so small compared to 1, 
which is the case in the endpoint region for the lepton (photon) energy spectrum 
in $B \rightarrow X_{u} \ell \bar{\nu}$ 
($B \rightarrow X_{s}\gamma$),  
the higher-order power corrections $(\Lambda_{\rm QCD} t)^n$ are 
important\footnote{For the regions with $m_{b} \gg 1/t$,
these power corrections are {\it enhanced} compared with the usual subleading corrections,
which are suppressed by powers of $\Lambda_{\rm QCD}/m_{b}$~\cite{Bauer:2001mh}.
Physically, this corresponds to the situation in which the decay spectra in the endpoint region 
is smeared over a range $\Delta$ with $m_{b} \gg \Delta \gtrsim \Lambda_{\rm QCD}$ \cite{Bauer:2003pi,ne}.}
to determine the behavior of $\tilde{f}(t)$.
Among the parameters (\ref{eq:momek})-(\ref{eq:momey}) involved in  ${\cal A}_{n}$, 
the experimental value is known for ${\cal K}_{0}$ $(=-\lambda_{1})$
of (\ref{eq:mnt}) only~\cite{Cronin-Hennessy:2001fk,blnp}.
Hopefully, further information regarding these generalized HQET parameters will be obtained from 
nonperturbative calculations
with, e.g., lattice QCD\footnote{
{}For recent progress in treating 
matrix elements of $B$-mesons in lattice QCD, see, e.g., Ref.~\citen{Onogi:rs}.}
and QCD sum rules.

\section{Fermi motion and four-parton correlation effects}

In this section we derive an explicit formula 
for the momentum representation of the shape function,
which is exact up to perturbative ${\cal O}(\alpha_{s})$ corrections.
Although this formula suffers from (large) perturbative corrections,
it is useful for obtaining some insight into the relevant nonperturbative effects,  
the Fermi motion of the $b$-quark, and 
the four-parton correlation effects,
without recourse to the $t$-expansion as in (\ref{eq:power}).
For this purpose, we derive the Fourier transformation of (\ref{eq:sol1})
into the residual momentum representation [see (\ref{eq:sf0})]. 
We obtain 
$f(\omega) = \delta(\omega) + [\mbox{contribution from } \tilde{J}]$.
Clearly, the first term, $\delta(\omega)$, reflects the lack of heavy-quark motion
inside the $B$-mesons in the ``non-interacting limit'',
and the above decomposition of 
$f(\omega)$
into the ``free'' term $\delta(\omega)$ and the ``interaction-dependent'' contribution from $\tilde{J}$
would 
be unsuitable 
to treat the nonperturbative part in the shape function.
In order to avoid such an
``artificial''
singularity, 
we rearrange the terms in (\ref{eq:sol1}).
Actually, it is more convenient to make the corresponding rearrangement in our differential equation (\ref{eq:de1})
by moving a certain term from the ``source'' on the RHS to the LHS, and then solve the resulting equation
with the new source.   
We decompose $\tilde{K}(t)$ in (\ref{eq:jtilde}) as [see (\ref{eq:K})]
\be
\tilde{K}(t) = 
\sigma^{2} \tilde{f}(t) + \delta \tilde{K}(t)\ ,
\label{eq:dK}
\ee
and move the first term with the constant $\sigma^2$
to the LHS of (\ref{eq:de1}). 
The residual term $\delta \tilde{K}(t)$ is treated as a source term.
The explicit form of the residual term reads
\be
\delta \tilde{K}(t) = \langle \bar{B}(v) | \bar{h}_{v}(z)
\left( (\stackrel{\leftarrow}{D_{\perp}}_{\mu} )^{2} 
- \sigma^{2} \right) h_{v}(0)| \bar{B}(v) \rangle\ .
\label{eq:dK2}
\ee
The rearrangement using (\ref{eq:dK}) is motivated by the observation that 
the first term 
is special among the terms on the RHS of (\ref{eq:jtilde}):
Among the nonlocal operators of (\ref{eq:K})-(\ref{eq:ty})
contributing to (\ref{eq:jtilde}),
the operator for $\tilde{K}(t)$ has the lowest mass dimension, and
it is expected to play a dominant role.
In fact, only this contribution is of a bilocal nature as (\ref{eq:K}), as in the case of
the shape-function operator (\ref{eq:sf0}), 
while the other contributions come from genuine many-body operators. 
Noting the properties of the higher twist nucleon structure functions in the DIS~\cite{Anthony:2002hy}, 
we conjecture that multi-parton correlation effects with increasing numbers of partons 
are less important. 
We note that the operator for $\tilde{K}(t)$ is a nonlocal version of
the kinetic energy operator of (\ref{eq:lambda1}), and it should be relevant to the Fermi motion of the $b$-quark
inside the $B$-meson;
in (\ref{eq:dK}) we have extracted the part proportional to $\tilde{f}(t)$ from $\tilde{K}(t)$ 
on the basis of the fact that $\tilde{K}(t)$ and $\tilde{f}(t)$ have similar bilocal natures,
and thus $\sigma$ would eventually correspond to the ``average measure''
of the transverse quark motion, as $\sigma^2 \sim - \lambda_{1}$
[note that $\tilde{K}(0) = - \lambda_{1}$].
However, for the time being, we can leave $\sigma$ as an arbitrary parameter.

In momentum space,
our differential equation now reads
\begin{equation}
\left(\omega^{2}-\sigma^{2} \right)\frac{df(\omega)}{d\omega}=-I(\omega)\ .
\label{eq:diffeqom}
\end{equation}
Here, the source $I(\omega)$, defined as
$\left.d\tilde{J}(t)/dt \right|_{\tilde{K}\rightarrow \delta\tilde{K}}\equiv 
i\int d\omega e^{i\omega t}I(\omega)$,
is given explicitly as [see (\ref{eq:jtilde})]
\begin{eqnarray}
I(\omega) &=& - \frac{d}{d\omega} \delta K(\omega)
+ \int_{-\infty}^{\omega}d\alpha\ {\rm P}\int_{-\infty}^{\infty}d\beta\
\frac{1}{\alpha - \beta}\left( \frac{\partial^2}{\partial \alpha^{2}}
+\frac{\partial^2}{\partial \beta^{2}}\right) W(\alpha, \beta)
\nonumber \\
&+& 
2\int_{-\infty}^{\omega}d\alpha\ {\rm P} \int_{-\infty}^{\infty}d\beta \int_{-\infty}^{\infty}d\xi\
\frac{1}{\alpha - \beta}\left\{
\left[ \frac{1}{\alpha-\xi}\left( \frac{\partial}{\partial \alpha}
+\frac{\partial}{\partial \xi}\right)^2
\right.\right.
\nonumber\\
&+&\frac{1}{\beta-\xi}\left( \frac{\partial}{\partial \beta}
+\frac{\partial}{\partial \xi}\right)^2
+ \frac{1}{(\alpha -\xi)^2}\left( \frac{\partial}{\partial \alpha}
+\frac{\partial}{\partial \xi}\right)
\nonumber\\
&+&\left.\left.  \frac{1}{(\alpha -\xi)(\beta-\xi)}\left( \frac{\partial}{\partial \beta}
+\frac{\partial}{\partial \xi}\right)\right]Y(\alpha, \beta, \xi)
- \frac{1}{\xi^2}\frac{\partial}{\partial \alpha}
Y(\alpha -\xi, \beta, \alpha)\right\}\ ,
\label{eq:iy}
\end{eqnarray}
where ``${\rm P}$'' denotes the principal value, and we have 
introduced the momentum representation for the relevant matrix elements
(\ref{eq:dK2}), (\ref{eq:tw}), and (\ref{eq:ty}) as
\begin{eqnarray}
\delta \tilde{K}(t) &=&\int d\omega\ e^{i\omega t}\delta K(\omega)\ , \label{eq:komega}\\
\tilde{W}(t, u) &=&\int d\omega\ d\omega' e^{i\omega' t-i(\omega'-\omega)ut}W(\omega, \omega')\ , 
\label{eq:womega}\\
\tilde{Y}(t, u, s) &=& \int  d\omega d\omega'd\xi\ e^{i\omega' t-i(\omega'-\xi)ut-i(\xi-\omega) st}
Y(\omega, \omega', \xi)\ . \label{eq:yomega}
\end{eqnarray}
We note that the parity transformation combined with the time-reversal
transformation
implies the symmetry properties
\begin{equation}
W(\omega,\omega') = W(\omega', \omega), \;\;\;\;\;\;\;\;\;\;\;\;\;\;  
Y(\omega, \omega', \xi)=Y(\omega',\omega, \xi)\ ,
\label{eq:symmetry}
\end{equation}
and hermiticity guarantees that $W(\omega, \omega')$ and $Y(\omega, \omega', \xi)$
are real functions. For both functions, the variable $\omega$ has the physical meaning of the 
residual light-cone momentum carried by the $b$-quark.
The quantity $\omega'-\omega$ for $W(\omega, \omega')$ represents the light-cone momentum 
carried by the gluon or the light $q\bar{q}$-pair, while  $\xi-\omega$ and $\omega'-\xi$
for $Y(\omega, \omega', \xi)$ represent the light-cone momenta carried by the gluons.
By inserting a complete set of states between the constituent fields,
we also get the support property: $W(\omega, \omega')$, as well as $Y(\omega, \omega', \xi)$,
vanishes unless $\omega < \bar{\Lambda}$ and $\omega' < \bar{\Lambda}$.
Similarly, we deduce that $\delta K(\omega)$ is a real function that vanishes unless $\omega < \bar{\Lambda}$.
Using these symmetry and support properties of $\delta K, W$ and $Y$,
it is straightforward to see that
$I(\omega)$ of (\ref{eq:iy})
vanishes unless $\omega < \bar{\Lambda}$.

We solve (\ref{eq:diffeqom}) with the boundary conditions $f(\omega)=0$ for $\omega \rightarrow \pm \infty$
and the normalization condition $\int d\omega f(\omega) = \tilde{f}(0) =1$
in the form
\be 
f(\omega) = f^{(0)}(\omega) + f^{(1)}(\omega)\ ,
\label{eq:soldec}
\ee
where $f^{(0)}(\omega)$
is the solution with the source $I(\omega)$
set to zero, while 
$f^{(1)}(\omega)$ denotes the piece induced by $I(\omega)$.
It is straightforward to obtain the analytic solution 
\be
f^{(0)}(\omega)= \frac{1}{2 \sigma}
\left\{ \theta(\omega+\sigma)
- \theta(\omega-\sigma)\right\} \ ,
\label{eq:solf0m}
\ee
which satisfies $\int d\omega f^{(0)}(\omega)=1$.
Also, the solution for $f^{(1)}(\omega)$ reads
\begin{equation}
f^{(1)}(\omega) = \frac{1}{2\sigma}\left(
\int_{\omega}^{\infty(\omega-\sigma)}d\omega'\ 
\frac{I(\omega')}{\omega' - \sigma}
- \int_{\omega}^{\infty(\omega + \sigma)}d\omega'\
\frac{I(\omega')}{\omega' + \sigma}\right)\ ,
\label{eq:solsol}
\end{equation}
which satisfies $\int d\omega f^{(1)}(\omega) =0$. 
The solution (\ref{eq:soldec}) with (\ref{eq:solf0m}) and (\ref{eq:solsol})
represents an explicit formula for the 
shape function, 
which satisfies the  
HQET equations of motion exactly.
We note that, if we take the limit $\sigma \rightarrow 0$,
we get $f^{(0)}(\omega) \rightarrow \delta(\omega)$, so that 
(\ref{eq:solf0m}) and (\ref{eq:solsol})
would reduce to the ``singular'' decomposition of $f(\omega)$ without the rearrangement using 
(\ref{eq:dK}).
Therefore, (\ref{eq:solf0m}) involves a nonperturbative effect represented by the parameter $\sigma$,
which smears the singular behavior of $\delta(\omega)$.
If we choose, e.g., $\sigma = \sqrt{- \lambda_{1}}$, the corresponding effect would be interpreted as 
the Fermi motion effect
of the $b$-quark inside the $B$-meson. This choice can be useful in practice, because, in the coordinate representation,
we have
\begin{equation}
\left.\tilde{f}^{(0)}(t)\right|_{\sigma = \sqrt{- \lambda_{1}}} 
= \frac{\sin\left( \sqrt{- \lambda_{1}}\ t\right)}{\sqrt{- \lambda_{1}}\ t}
=1+\sum_{k=1}^{\infty}\frac{t^{2k}\lambda_{1}^{k}}{(2k)!(2k+1)}\ ,
\label{eq:exp0}
\end{equation}
and the expansion on the RHS reproduces the first three terms of (\ref{eq:power}) exactly [see (\ref{eq:mnt}),
and note that 
$\left.\tilde{f}^{(1)}(t)\right|_{\sigma = \sqrt{- \lambda_{1}}} =it^3{\cal W}_{0,0}/36+
{\cal O}(t^4)$].
In fact, ``flat distribution'' (\ref{eq:solf0m}) with $\sigma = \sqrt{- \lambda_{1}}$
exhibits behavior consistent with the Gaussian distribution around $\omega=0$ with the width $\sqrt{-\lambda_{1}}$, 
which was proposed in Refs.~\citen{Korchemsky:1994jb} and \citen{Grozin:1994ni}
as a nonperturbative ansatz valid in the vicinity of $\omega=0$.
As an alternative to the choice $\sigma = \sqrt{- \lambda_{1}}$, one may employ another ``optimized'' choice for $\sigma$,
which makes $\delta \tilde{K}(t)$ of (\ref{eq:dK2})
``small'', so that $\delta \tilde{K}(t)$ could be treated as a perturbation.

It is well known that (\ref{eq:sf0})
should obey
$f(\omega)=0$ for $\omega >\bar{\Lambda}$~\cite{Neubert:1993um,Bigi:1993ex,Mannel:1994pm}.
Using the above-mentioned support property for $I(\omega)$, 
it is straightforward to show that our solution (\ref{eq:soldec}) indeed
vanishes unless $\omega < \bar{\Lambda}$,\footnote{This is actually satisfied even in 
the case $\sigma > \bar{\Lambda}$,
as well as for $\sigma \le \bar{\Lambda}$.}
reproducing the correct support property.

Our solution reveals that the shape function is subject to nonperturbative effects 
due to the multi-parton correlation
with additional quarks and gluons, which are 
represented by $W(\omega, \omega')$ and $Y(\omega,\omega', \xi)$,
as in (\ref{eq:solsol}), (\ref{eq:iy}).
This indicates that 
$f(\omega)$
is actually a complicated multi-particle object beyond a
``simple'' momentum distribution function. 
This is in contrast to the leading-twist nucleon structure functions in the DIS,
but it is reminiscent of the character of the higher-twist ones \cite{Balitsky:1987bk,Ali:1991em,Kodaira:1998jn}.
Our results, (\ref{eq:solsol}) and (\ref{eq:iy}), are exact to lowest order in the perturbative effects, 
but the multi-particle character of $f(\omega)$ should persist when perturbative corrections are included.

\section{Conclusion}

The shape function of the $B$-meson is an important 
ingredient in the factorization formula of the differential rates 
of inclusive $B$-meson decays, such as $B\rightarrow X_s\gamma$ and 
$B\rightarrow X_u l\bar{\nu}$. 
We studied the ``soft components'' of the shape function 
separately from the ``hard (perturbative) components'' relevant to its scale dependence,
and we studied the model-independent constraints on the soft components in the heavy quark limit.
From the 
equations of motion and heavy-quark symmetry constraints, 
we derived a differential 
equation that relates the longitudinal-momentum dependence of the shape function 
to matrix elements of the novel nonlocal operators. 
Solving this equation, we obtained a new basis of nonlocal operators, 
which describes
the relevant soft components of the shape function 
in terms of the kinetic energy distribution and the four-parton correlations.
We also derived 
the relations among the matrix 
elements of local operators with arbitrary numbers of covariant derivatives.
These relations not only reproduce all the known results for 
a few covariant derivatives
but also include new relations for more covariant derivatives.
These relations are exact only when we ignore the perturbative effects,
but they
are useful beyond the lowest order in the perturbative effects 
for the purpose of finding a basis of independent local operators with the same dimension.
Our relations indicate that the local operator basis is composed of 
an increasing number of operators with increasing dimensions
whose matrix elements give
the generalized HQET parameters that
represent the Fermi motion of the $b$-quark as well as the four-parton correlation effects
inside the $B$-meson.
Our differential equation also yields
the momentum representation $f(\omega)$ 
as a sum of the part $f^{(0)}(\omega)$ involving the Fermi motion
and the additional 
part $f^{(1)}(\omega)$ induced by more sophisticated correlation effects.
The behavior of the ``leading'' part, $f^{(0)}(\omega)$, appears to be
consistent with the existing nonperturbative ansatz~\cite{Korchemsky:1994jb,Grozin:1994ni} 
for the shape function. 
Our momentum-representation formula is again exact up to perturbative corrections,
but it provides insight into model building of the relevant nonperturbative effects.

Our results 
reveal that, due to nonperturbative effects, 
$f(\omega)$ is a much more complicated object
than the simple momentum distribution of the $b$-quark inside the $B$-meson:
Nonperturbative effects induce the mixing of 
additional dynamical quarks and gluons, so that the 
shape function contains the soft components that represent the multi-particle correlation effects.
In connection to this, it is interesting to note that
a probabilistic interpretation of $f(\omega)$ as the momentum distribution
has been questioned from a different point of view, noting
that $\int d\omega f(\omega)$ is negatively divergent, due to
the {\it perturbative} effects in (\ref{eq:sf0}), as mentioned in \S1~\cite{Bauer:2003pi,blnp}.

The ``full'' shape function (\ref{eq:sf0}) entering into the factorization formula for the inclusive decay rates
is obtained by combining the soft components with the hard components
through the matching procedure.
We emphasize that 
(\ref{eq:basis}) with (\ref{eq:momek})-(\ref{eq:momey})
provides a 
local operator basis that is
necessary for parameterizing 
the soft components in the matching calculation. Furthermore, 
(\ref{eq:power}) with (\ref{eq:momsol}) completes the matching at the tree level
for operators of arbitrary dimension,
and the corresponding leading-order matching coefficients 
can be easily read off from this formula.
One can proceed to the one-loop matching calculations using 
our 
operator basis, 
and the loop effects will produce the 
mixing coefficients at the next-to-leading order 
between the operators with the same dimension.
Also, the nonlocal operator basis provided by (\ref{eq:K})-(\ref{eq:ty}) may be useful,
and (\ref{eq:sol1}) ((\ref{eq:soldec}) with (\ref{eq:solf0m}) and (\ref{eq:solsol})) completes
the corresponding leading-order matching in the coordinate (momentum) space.
Using the results of the matching calculations, a detailed study of the interplay 
between the strong perturbative effects in the hard components,
which generate a Sudakov-type scale dependence,
and the nonperturbative effects in the soft components, which have been unraveled in this paper,
should clarify the behavior of the full shape function and its roles in decay rates.
This will be studied in a separate publication.


\section*{Acknowledgements}
The authors would like to thank T. Onogi and S. Hashimoto for discussions
that stimulated this investigation.
The work of J. K. is supported in part by the 
Grant-in-Aid
for Scientific Research (No. C-16540255).
The work of K. T. is supported in part by the Sumitomo Foundation
and by the 
Grant-in-Aid
for Scientific Research (No. C-16540266).

\end{document}